\documentclass[conference]{IEEEtran}
\IEEEoverridecommandlockouts

\newcommand{\red}[1]{\textcolor{black}{#1}} 

\usepackage{amsmath}
\usepackage{algorithm}

\usepackage{cite}
\usepackage{bm}
\usepackage{amsmath,amssymb,amsfonts}
\usepackage{algorithmic}
\usepackage{graphicx}
\usepackage{textcomp}
\usepackage{xcolor}

\usepackage{draftwatermark}
\SetWatermarkText{ACCEPTED \\MANUSCRIPT}
\SetWatermarkScale{.6}
\SetWatermarkColor[gray]{0.9}

\begin{document}

\title{Improving Network Anomaly Detection via Choquet-Integral-Based Feature Aggregation\thanks{This manuscript has been accepted for presentation at the IEEE International Symposium on Computers and Communications (ISCC 2026).}
}

\author{\IEEEauthorblockN{ 
Abreu Quevedo\IEEEauthorrefmark{1},
Roger Immich\IEEEauthorrefmark{2},
Giancarlo Lucca\IEEEauthorrefmark{1},
Gra\c{c}aliz Dimuro\IEEEauthorrefmark{1},
Bruno L. Dalmazo\IEEEauthorrefmark{1}
} 
\IEEEauthorblockA{\IEEEauthorrefmark{1}Federal University of Rio Grande - FURG, Brazil}%
\IEEEauthorblockA{\IEEEauthorrefmark{2}Federal University of Rio Grande do Norte - UFRN, Brazil}%
E-mail: \{abreu\_rg, giancarlo.lucca, gracalizdimuro, dalmazo\}@furg.br, roger@imd.ufrn.br
}


\maketitle

\begin{abstract}

This work investigates a generalized Choquet-integral-based feature aggregation framework to improve anomaly detection in high-dimensional network traffic data. The approach combines adaptive weighting with incremental feature selection to address feature redundancy. Using Random Forest and XGBoost classifiers, we evaluate models trained with both raw and Choquet-aggregated features under varying feature subset sizes. The proposed aggregation achieves up to $7\%$ higher accuracy while reducing data volume by $77.5\%$ (from $214$~MB to $48$~MB), without degrading precision and recall. Results averaged over multiple stratified repetitions indicate that Choquet-based aggregation yields statistically significant gains ($p < 0.05$) in scenarios with limited feature availability, highlighting its suitability for real-time intrusion detection under bandwidth and feature-availability constraints.

\end{abstract}


\begin{IEEEkeywords}
Choquet Integral, Feature Engineering, Anomaly detection, Fuzzy.
\end{IEEEkeywords}

\section{Introduction}

In recent decades, distributed computing infrastructures have been the cornerstone of human development, especially in the field of communication. Every day, we become more accustomed to dealing with increasingly complex network patterns that continue to grow over time. Considering the current society in which we live, where a significant portion of humanity constantly generates data, depending on network management, IoT services~\cite{ISCC_Citation} and communication within \red{ Smart City ecosystems}, there is an increasing dependence on ensuring the reliability of this information. This context highlights critical aspects related to its vulnerability \cite{dalmazo2018} and security \cite{quevedoRfXG} within contemporary digital networks. 

In this context, the massive flow of information in urban environments  plays a crucial role not only in communication but also in maintaining security. Nevertheless, challenges such as data leakage, exposure of user information, and authentication vulnerabilities persist in the current digital landscape. Moreover, the increasing complexity of network patterns demands more sophisticated mechanisms for identifying subtle anomalies and for processing and extracting meaningful insights~\cite{CARDOSO2024111639} \red{especially where bandwidth constraints and real-time processing are critical.}

DDoS attacks, or Distributed Denial-of-Service attacks, follow the same evolutionary pattern as data itself, becoming increasingly powerful and large-scale over time.
Some examples of the magnitude of these attacks were highlighted in 2025, when Cloudflare mitigated a record breaking distributed denial-of-service (DDoS) attack that peaked at 29,7 Tbps\cite{cloudflareRef}. Such events illustrate the unprecedented scale of modern DDoS threats and how it  \red{reinforces the urgency of improving anomaly detection methods, particularly to safeguard the critical communication backbones of modern Smart Cities.} In 2024, Cloudflare’s autonomous DDoS defense systems blocked approximately 21.3 million DDoS attacks, marking a 53\% increase compared to 2023, on average, the company mitigated 4870 DDoS attacks per hour throughout the year and in the same year the company reported an expressive 1885\% increase in the number of attacks above 1 Tbps in the last quarter of 2024, highlighting the severity and timeliness of this threat. A notable case in this context was the massive cyberattack against Elon Musk’s platform X, attributed to the hacker group Dark Storm, which caused a large-scale outage in March 2025~\cite{decrypt2025}.

Considering the importance of defending large-scale data traffic, as previously mentioned, the Choquet integral~\cite{cac_artigo,choquet19531954,ChoquetMed, test-amo} has emerged as a promising alternative across several data-intensive domains. In this sense, recent studies~\cite{quevedo2025choquet,quevedoRfXG, sbsegleite} have demonstrated its effectiveness in mitigating network-based attacks, highlighting its potential to enhance the robustness and adaptability of anomaly detection systems.
Following this perspective, we investigated how integrating fuzzy aggregation mechanisms, such as the Choquet integral, can favor a more expressive representation of inter-feature dependencies. This capability enables models to capture subtle correlations often overlooked by traditional techniques, contributing to a more accurate and resilient defense against evolving cyber threats.

This study analyzes feature aggregation using feature engineering based on the generalized Choquet integral. The experiment incrementally increases the number of features to observe the classification model's performance. Each setup is evaluated in two conditions: with features aggregated via the generalized Choquet integral and without aggregation. Random Forest and XGBoost are used to compare scenarios, assessing the impact of Choquet-based feature fusion on accuracy, precision, recall, and F1-score. This evaluation clarifies how fuzzy aggregation influences model performance as the feature space grows.

The rest of this paper is structured as follows. Section 2 reviews the related studies. Section 3 details the proposed workflow. Section 4 describes the implementation, experimental setup, and results. Finally, Section 5 discusses the conclusions and outlines possible directions for future research.

\section{Related Work}
In this section, we discuss research efforts relevant to the scope of this paper. Accordingly, studies that have made notable contributions to fuzzy modeling, dimensionality reduction, and anomaly detection systems were reviewed.

The study \cite{quevedo2025choquet} presents a network traffic predictor based on generalized Choquet integrals with adaptive weighting. Its goal is to improve large-scale traffic forecasting by analyzing how the parameter $\alpha$ affects both prediction accuracy and execution time. The authors used a processed version of the CIC-DDoS2019 dataset, applying sliding-window and aggregation techniques to capture temporal behavior. Three strategies were tested to select the optimized $\alpha$: brute force, binary search, and random binary search. The results show that the binary and random approaches considerably reduce execution time, achieving reductions of 56.75\% and 59.49\% compared to brute force, while maintaining good accuracy. Among the evaluated models, the Choquet (a) formulation achieved the lowest average prediction error. However, the study focuses only on traffic volume prediction and $\alpha$ selection, without extending the analysis to other network indicators or to improving classification-based detection methods, which is the focus of this work.

The study \cite{quevedoRfXG} investigates the use of the generalized Choquet integral as a feature engineering technique to improve the performance of Random Forest and XGBoost models. Using features from the CIC-DDoS2019 dataset, the authors evaluated classification performance under different aggregation settings. The results show clear gains for both models after introducing the Choquet-based feature. Random Forest achieved improvements of 4.59\% in accuracy, 37\% in recall, and 16\% in F1-score, while XGBoost showed gains of 4.46\% in accuracy, 35\% in recall, and 16\% in F1-score. Despite these positive results, the approach applied the Choquet integral only to the top-ranked feature and restricted the classification to four attributes, which differs from the broader analysis conducted in this work.

In \cite{LongFuzzyAnomaly} presents a modular anomaly detection and mitigation system for SDN environments. Based on Long Short-Term Memory with fuzzy logic (LSTM-FUZZY), the approach comprises three phases: characterization, detection, and mitigation. Testing involved SDN flows from Mininet/Floodlight and the CICDDoS 2019 dataset, validating the modules' effectiveness by operating autonomously and using the system to eliminate human intervention and supports network administrators in maintaining operational efficiency. Its modular architecture is a key strength, allowing the integration of diverse techniques an essential feature for adapting to evolving network dynamics and security demands.

\section{Proposal}
This study proposes a systematic evaluation framework to examine how feature-space growth impacts classifier performance when features are fused via the generalized Choquet integral versus a non-aggregated baseline. Concretely, we construct incremental feature sets of size 
k=1,…,10 and evaluate each configuration under two parallel pipelines: (i) Choquet-based feature fusion and (ii) the same features without Choquet aggregation. We then train Random Forest and XGBoost models on both pipelines using identical data splits and report accuracy, precision, recall, and F1-score. By comparing performance trajectories as k increases and by analyzing the deltas between the pipelines, this research aims to quantify the contribution of fuzzy aggregation to predictive effectiveness as the feature space expands. In this work, anomaly detection is treated in the context of smart cities, where cognitive clouds must process massive streams of data from diverse IoT sensors. \red{Managing high-dimensional traffic becomes a critical challenge for privacy-aware and latency-sensitive applications.}

\subsection{Conceptual Model}
\begin{figure}[htbp]
    \centering
\includegraphics[width=0.8\columnwidth]{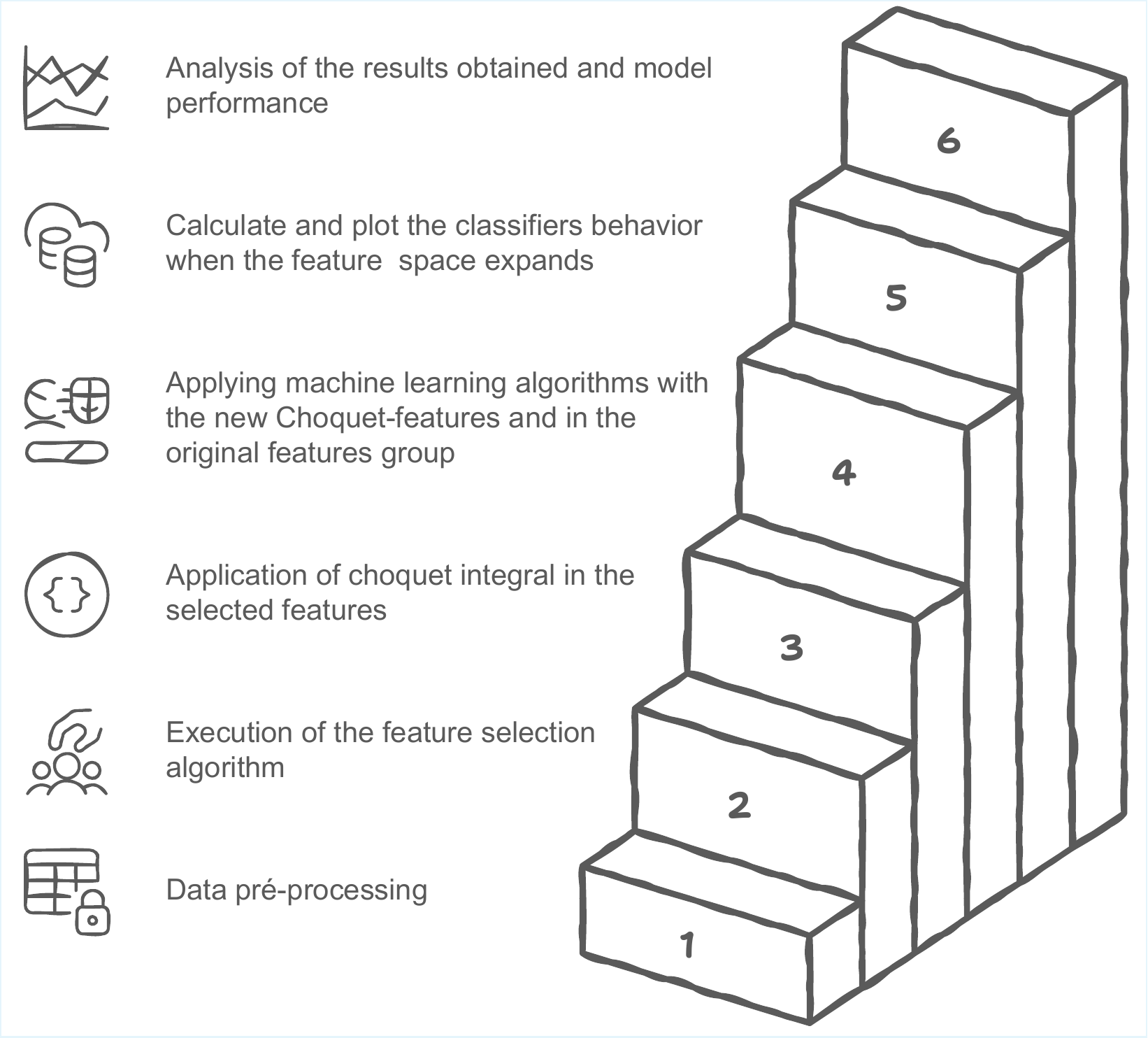}
    \caption{Conceptual workflow illustrating the stages from data preprocessing to model evaluation, including feature selection and Choquet-based feature construction.}
    \label{fig1}
\end{figure}

Figure~\ref{fig1} serves as the conceptual basis of this study, outlining its main stages and illustrating the workflow, key components, and their interrelations.

\begin{enumerate}

    \item \textbf{Data pre-processing}: 
    The network dataset is cleaned, normalized, and organized into a structured format suitable for the generalized Choquet Integral algorithm, ensuring data consistency for subsequent phases. (Step~1 in Fig.~\ref{fig1})

    \item \textbf{Execution of the feature selection algorithm}: 
    A feature selection procedure ranks attributes based on their statistical contribution to the classification task, identifying the most informative subset for the model. (Step~2 in Fig.~\ref{fig1})

    \item \textbf{Application of Choquet integral in the selected features}: 
    Highest-ranked features are progressively aggregated via the generalized Choquet integral to create synthetic attributes that capture nonlinear interactions and dependencies. (Step~3 in Fig.~\ref{fig1})

    \item \textbf{Applying machine learning algorithms}: 
    Classifiers are trained using both the original feature set and the enhanced dataset (including Choquet-aggregated features) to enable a controlled comparison of both scenarios. (Step~4 in Fig.~\ref{fig1})

    \item \textbf{Calculate and plot classifier behavior}: 
    The feature space is incremented gradually ($k=1, 2, \dots, 10$), recording metrics such as accuracy, precision, recall, and F1-score to visualize performance evolution as dimensionality increases. (Step~5 in Fig.~\ref{fig1})

    \item \textbf{Analysis of results and performance}: 
    The collected metrics are examined to quantify the influence of Choquet-based aggregation on predictive capacity, highlighting the approach's impact on classification performance. (Step~6 in Fig.~\ref{fig1})

\end{enumerate}

\subsection{Aggregation Functions and the Choquet Integral}

The Choquet integral~\cite{axioms12040330} is a central aggregation operator in information fusion and decision-making problems. Let $\mathfrak{m}\colon 2^{N} \rightarrow [0,1]$ be a fuzzy measure defined on the power set of $N=\{1,\ldots,n\}$~\cite{FuzzyMeasures:Murofushi}. The discrete Choquet integral associated with $\mathfrak{m}$ is the mapping $\mathfrak{C}_m\colon [0,1]^n \rightarrow [0,1]$ given, for any $\bm{x} \in [0,1]^n$, by
{\small
\begin{equation}\label{eq:intChoquet}
	\mathfrak{C}_m(\bm{x}) = \sum_{i=1}^n \left( x_{(i)} - x_{(i-1)} \right) m\!\left( A_{(i)} \right),
\end{equation}
}
where $(x_{(1)}, \ldots, x_{(n)})$ denotes an increasing permutation of the components of $\bm{x}$, with $x_{(0)} = 0$, and $A_{(i)} = \{(i), \ldots, (n)\}$ represents the subset of indices corresponding to the $n-i+1$ largest components of $\bm{x}$.

By distributing the product in~\eqref{eq:intChoquet}, the Choquet integral can be equivalently expressed in its expanded form as
{\small\begin{equation}\label{eq:intChoquet2}
    \mathfrak{C}_{m}(\bm{x}) = \sum_{i=1}^{n} \left( x_{(i)} \, m\!\left(A_{(i)} \right) - x_{(i-1)} \, m\!\left(A_{(i)} \right) \right).
\end{equation}}

This expanded representation serves as the foundation for the CC-integral~\cite{ChoquetMed}, a generalization of the Choquet integral in which the product operator is replaced by a copula function~\cite{alsina2006associative}. 

Let the operator $m \colon 2^{N} \rightarrow [0,1]$ be a fuzzy measure and consider a bivariate copula $C \colon  [0,1]^2 \rightarrow [0,1]$. The CC-integral is defined as a mapping $\mathfrak{C}_{m}^C \colon [0,1]^n \rightarrow [0,1]$, given,  $\forall {\bm{x}=(x_1,\ldots, x_n)} \in [0,1]^n$, by
	{\small \begin{equation}\label{eq:intChoquet_extCCintegral}
		\mathfrak{C}_{m}^C ({\bm x}) = \sum_{i=1}^{n} C \left ( x_{(i)}, m\left(A_{(i)} \right)\right ) - C \left ( x_{(i-1)}, m\left(A_{(i)} \right) \right ),
    \end{equation} }
where $ x_{(i)} $ and $A_{(i)}$ are defined in the same way as Eq.~(\ref{eq:intChoquet}).

Building upon this idea, a parametric family known as the $C_{\alpha}C$-integrals s~\cite{ChoquetMed} was introduced by combining CC-integrals with $\alpha$-parameterized copulae, like the ones shown in Table~\ref{tabela_exemplo_copulas}.  The $C_{\alpha}C$-integrals constitute the aggregation framework adopted in this work. 

{The {Poisson Moving Average} (PMA) is employed as a weighting factor and used as fuzzy measure, being defined from a truncated Poisson distribution parameterized by the sliding window size \(W\). The weights associated with each position \(i\) within the window are given by
{\small\begin{equation}
\mu_i = \frac{W^i e^{-W}}{i!}, \quad i = 1, 2, \ldots, W,
\end{equation}}
and are subsequently normalized such that 
sum from i = 1 to $W$ of $\mu_i$ = 1.
This approach allows adjusting the temporal influence of the elements within the window, making the aggregation process more sensitive to the dynamics of the analyzed data.}

\begin{table*}[htbp]
    \centering
    \setlength{\tabcolsep}{3pt} 
    \small 
    \caption{Table of Generalizations of the Choquet Integral Used to Generate a New Feature.}
    \label{tabela_exemplo_copulas}
    \begin{tabular}{l c c c}
        \hline
        \textbf{Copula} & \textbf{Functions} & \textbf{$\alpha$ constraints} \\
        \hline 
        (A) & $C_\alpha (x,y) = xy[1 + \alpha (1-x)(1-y)]$ & $-1 \leq \alpha \leq 1 $\\ 
        (B) & $C_\alpha (x,y) = \frac{1}{1 + \alpha} \max[x+y-1+\alpha - \alpha |x-y|,0]$ & $0 < \alpha <1$  \\
        (C) & $C_\alpha = (1 - \alpha)W + \alpha \min $ & $0 < \alpha <1$  \\
        (D) & $C_\alpha = \frac{\alpha ^{2} (1-\alpha)}{2} W + (1 - \alpha ^{2}) P + \frac{\alpha ^{2} (1+\alpha)}{2} \min$ & $-1 < \alpha < 1 (\alpha \neq 0)$ \\
        \hline 
    \end{tabular}
\end{table*}

\subsection{Work definitions}
Before presenting the experimental results, it is necessary to clarify some concepts used throughout this work. 

\subsubsection{Choquet-based feature}

In this work, the term \textbf{Choquet-based feature} refers to a new synthetic attribute generated through the generalized Choquet Integral. 
Another key aspect is the parameter $k$, which denotes the number of selected features used in each experimental configuration.

\subsubsection{$\Delta Y$ (Performance gain)}

Throughout this paper, the symbol $\Delta Y$ represents the improvement obtained when using the Choquet-based aggregated feature compared to the baseline without aggregation.

where $Y$ is the accuracy and a positive $\Delta Y$ indicates that the Choquet aggregation improves model performance, while a negative value indicates performance degradation.

This metric enables direct comparison between both approaches at each feature-set size $k$, allowing identification of the most effective operating range for the aggregated feature.

\subsection{Classify Algorithm Methods}

This study evaluates the proposed feature engineering approach using two tree-based ensemble methods: \textbf{Random Forest (RF)} \cite{breiman2001random} and \textbf{XGBoost} \cite{chen2016xgboost}. While both rely on decision tree architectures, they utilize fundamentally different construction strategies:

\begin{itemize}
    \item \textbf{Random Forest:} Employs a bagging technique where multiple trees are built independently using bootstrap samples and random feature subsets. The final classification is reached through majority voting, a process that minimizes correlation between trees, enhances generalization, and ensures robustness against noise in high-dimensional network data.
    \item \textbf{XGBoost:} Implements a gradient boosting framework where trees are trained sequentially to minimize the residual errors of the preceding ensemble. It incorporates $L1$ and $L2$ regularization and a specific learning rate to prevent overfitting, offering high predictive performance and efficiency in handling sparse datasets.
\end{itemize}

\subsubsection*{Comparison}

RF is characterized by its stability and reliable performance with minimal hyperparameter tuning, making it suitable for establishing a baseline in anomaly detection. In contrast, XGBoost typically achieves superior accuracy by capturing complex, non-linear patterns in traffic data, though it requires more precise tuning to balance performance and computational overhead.

\section{Evaluation}
To assess the effectiveness of the proposed feature engineering approach, this section analyzes how the feature space evolves as more features are included.

\subsection{Data preprocessing and feature selection algorithm}

Before generating the new Choquet–based feature, the dataset was preprocessed to ensure data consistency and model reliability. The raw CIC-DDoS2019 dataset composed of 692703 rows and 79 columns was cleaned using Pandas and NumPy, including handling missing values and converting categorical attributes into numerical form. After preprocessing, \red{a comparative analysis of different selection methods was conducted choosing SelectKBest, the algorithm was ultimately selected for providing the most consistent ranking for our classification task,} producing the top 10 most relevant attributes shown in Table~\ref{tab:selected_features}. This preprocessing and feature-selection also significantly reduces the dataset size by 77.5\%, shrinking it from 214~MB to 48~MB, which improves training efficiency and speeds up experimentation in bandwidth-constrained smart city environments.

The choice of selecting only the top 10 features was intentional. Preliminary experiments showed that, for both Random Forest and XGBoost, the performance tends to stabilize once more than 6–7 features are used as shown in the performance curves in Fig.~\ref{acRF} and \ref{acXG}, therefore, limiting the feature space to 10 attributes prevents the model from incorporating unnecessary information, while still allowing the evaluation of how performance evolves as the feature space gradually expands. This makes the analysis more controlled and avoids feeding excessive or redundant features to the classifiers.
\begin{table}[h]
\centering
\setlength{\tabcolsep}{3pt} 
\caption{Top-10 selected features based on SelectKBest.}
\label{tab:selected_features}
\begin{tabular}{|c|l|c|}
\hline
\textbf{Rank} & \textbf{Feature} & \textbf{Score} \\
\hline
1\textsuperscript{st}  & Bwd Packet Length Mean    & 112023.06 \\
\hline
2\textsuperscript{nd}  & Avg Bwd Segment Size      & 112023.06 \\
\hline
3\textsuperscript{rd}  & Bwd Packet Length Std     & 108310.46 \\
\hline
4\textsuperscript{th}  & Bwd Packet Length Max     & 106125.98 \\
\hline
5\textsuperscript{th}  & Packet Length Std         & 101429.54 \\
\hline
6\textsuperscript{th}  & Max Packet Length         & 94126.52  \\
\hline
7\textsuperscript{th}  & Fwd IAT Max               & 93922.66  \\
\hline
8\textsuperscript{th}  & Flow IAT Max              & 93525.75  \\
\hline
9\textsuperscript{th}  & Packet Length Mean        & 84086.98  \\
\hline
10\textsuperscript{th} & Packet Length Variance    & 82271.07  \\
\hline
\end{tabular}
\end{table}

\subsection{Accuracy Comparison: With vs. Without Choquet Feature}

\begin{table}[htbp]
    \centering
    \setlength{\tabcolsep}{3pt} 
    \caption{Accuracy difference ($\Delta Y$) between Choquet-based and baseline features for the first ten feature subsets (Random Forest).}
    \label{tab:dffRF_swapped}
    \begin{tabular}{|c|c|c|c|}
        \hline
        Feature &
        $Y_{\text{Choquet}}$ &
        $Y_{\text{Baseline}}$ &
        $\Delta Y = Y_{\text{Choquet}} - Y_{\text{Baseline}}$ \\
        \hline
        1  & 0.931 & 0.870 & 0.061 \\
        \hline
        2  & 0.934 & 0.870 & 0.064 \\
        \hline
        3  & 0.941 & 0.870 & 0.071 \\
        \hline
        \textbf{4} & \textbf{0.953} & \textbf{0.880} & \textbf{0.073} \\
        \hline
        5  & 0.948 & 0.909 & 0.039 \\
        \hline
        6  & 0.962 & 0.990 & -0.028 \\
        \hline
        7  & 0.964 & 0.984 & -0.020 \\
        \hline
        8  & 0.965 & 0.985 & -0.020 \\
        \hline
        9  & 0.964 & 0.984 & -0.020 \\
        \hline
        10 & 0.963 & 0.983 & -0.020 \\
        \hline
    \end{tabular}
\end{table}


\begin{table}[htbp]
    \centering
    \setlength{\tabcolsep}{3pt} 
    \caption{Accuracy difference ($\Delta Y$) between Choquet-based and baseline features for the first ten feature subsets (XGBoost).}
    \label{tab:dffXG}
    \begin{tabular}{|c|c|c|c|}
        \hline
        Feature &
        $Y_{\text{Choquet}}$ &
        $Y_{\text{Baseline}}$ &
        $\Delta Y = Y_{\text{Choquet}} - Y_{\text{Baseline}}$ \\
        \hline
        1  & 0.924 & 0.860 & 0.064 \\
        \hline
        2  & 0.923 & 0.857 & 0.066 \\
        \hline
        3  & 0.934 & 0.864 & 0.070 \\
        \hline
        \textbf{4}  & \textbf{0.951} & \textbf{0.880} & \textbf{0.071} \\
        \hline
        5  & 0.947 & 0.903 & 0.044 \\
        \hline
        6  & 0.962 & 0.986 & -0.024 \\
        \hline
        7  & 0.964 & 0.986 & -0.022 \\
        \hline
        8  & 0.965 & 0.986 & -0.021 \\
        \hline
        9  & 0.966 & 0.987 & -0.021 \\
        \hline
        10 & 0.967 & 0.988 & -0.021 \\
        \hline
    \end{tabular}
\end{table}

\begin{figure}[htbp]
    \centering
\includegraphics[width=0.8\columnwidth]{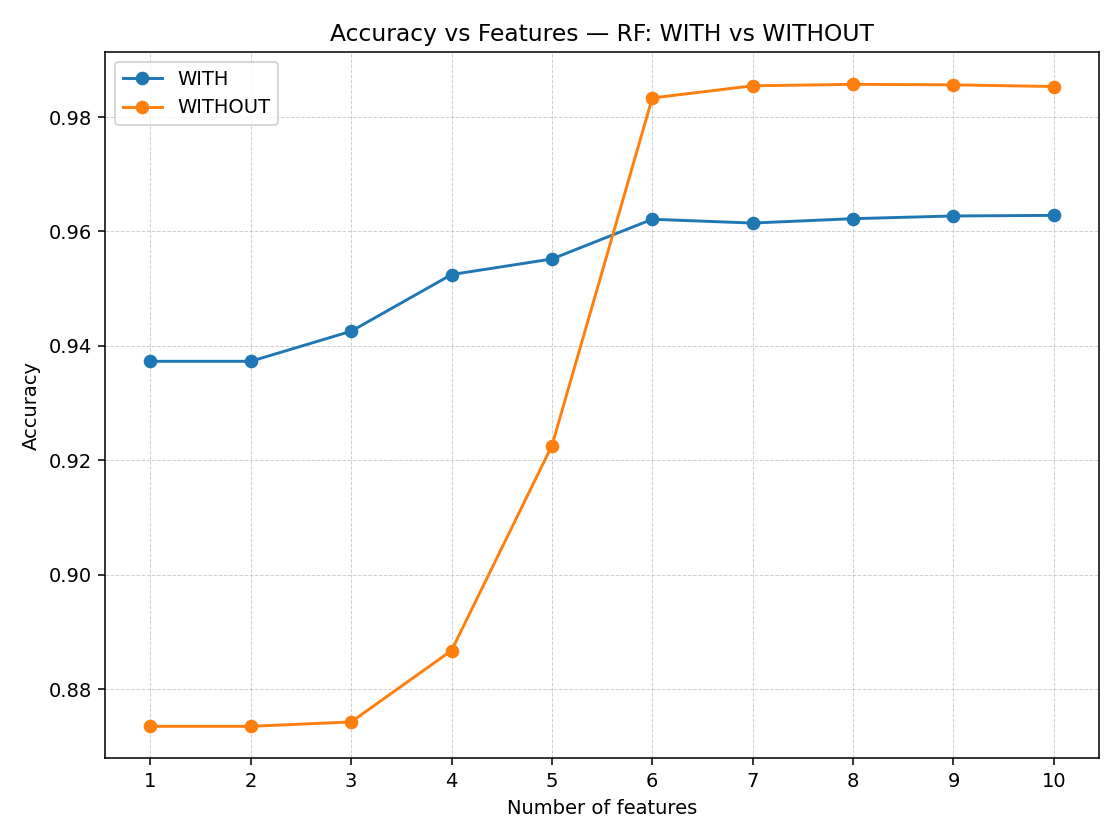}
    \caption{The evolution of the accuracy according to the number of features (Random Forest).}
    \label{acRF}
\end{figure}

\begin{figure}[htbp]
    \centering
\includegraphics[width=0.8\columnwidth]{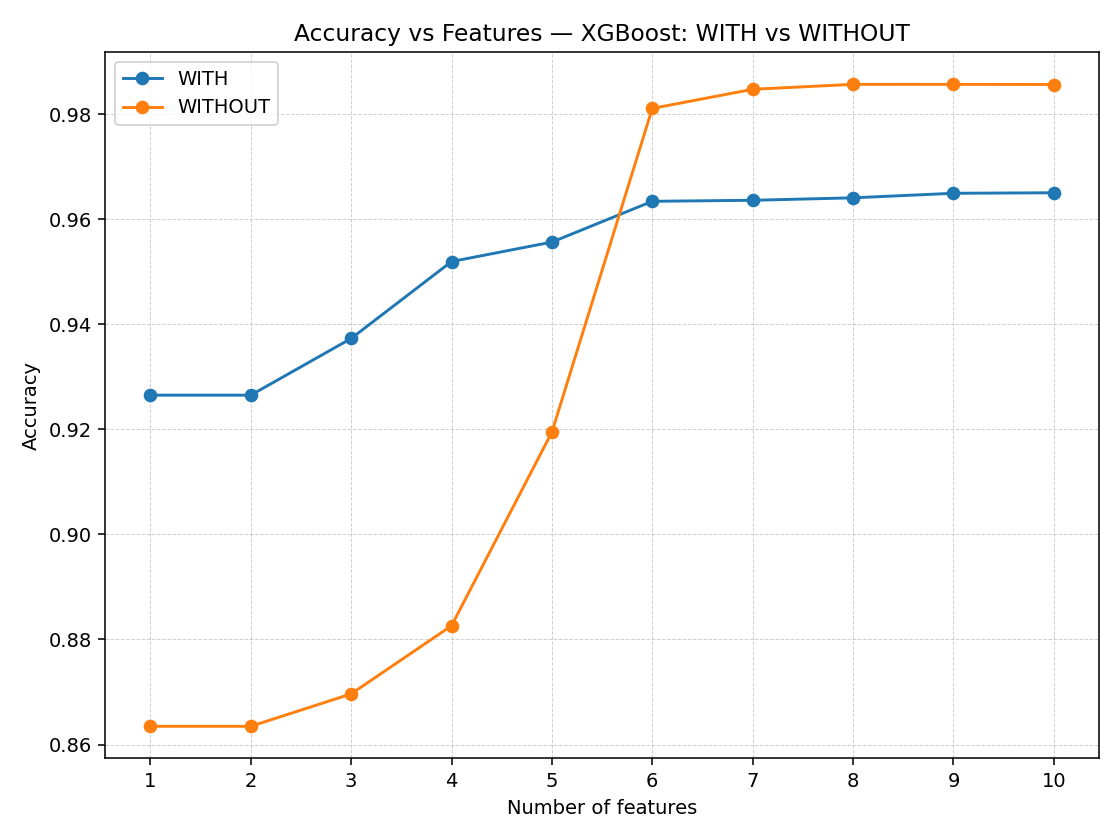}
    \caption{The evolution of the accuracy according to the number of features (XGBoost).}
    \label{acXG}
\end{figure}

Figs. \ref{acRF}, \ref{acXG} and Tables \ref{tab:dffRF_swapped}, \ref{tab:dffXG} present the accuracy evolution as the feature space grows (from 1 to 10 features), comparing the traditional feature set against the configuration that incorporates the Choquet-based feature.
As can be observed, when only a small number of features is available, the configuration with the Choquet–based feature yields consistently higher accuracy compared to the baseline without aggregation. As the number of features increases, the performance gap gradually narrows, and both approaches converge to similar accuracy levels once additional raw features are incorporated. The best $k$ for Random Forest and XGBoost, the largest accuracy gain occurs at $k=4$ ($\Delta Y=0.07$), indicating an effective positive operating window at $k\in[1,5]$ before both curves saturate and converge for $k \ge 6$.

\subsection{Detection performance}

To better understand how the Choquet-based feature impacts the classifiers,
confusion matrices were generated for both algorithms under two conditions:
(i) using only the raw selected features, and (ii) using the same features with the addition treatment of the Choquet-aggregated feature. Remember that 4 features are used for Random Forest and XGBoost, since Tables~\ref{tab:dffRF_swapped} and \ref{tab:dffXG} indicate that these configurations yield the highest accuracy for each model and so well the best ($\Delta Y$).

\begin{figure}[htbp]
\vspace{0.04in}
    \centering
\includegraphics[width=0.8\columnwidth]{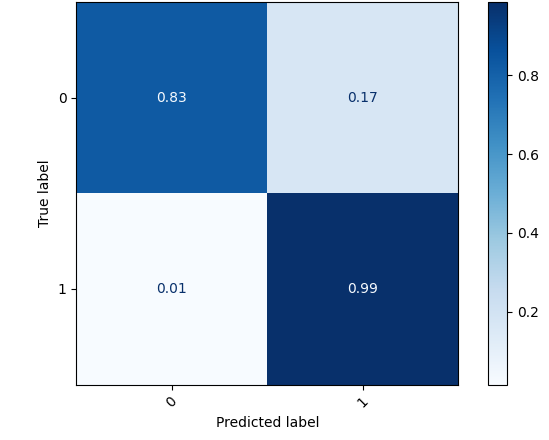}
    \caption{Confusion matrix of the classifier trained using only the original selected features (baseline) for 4 features.}
    \label{XGConfMatrix}
\end{figure}

\begin{figure}[htbp]
    \centering
\includegraphics[width=0.8\columnwidth]{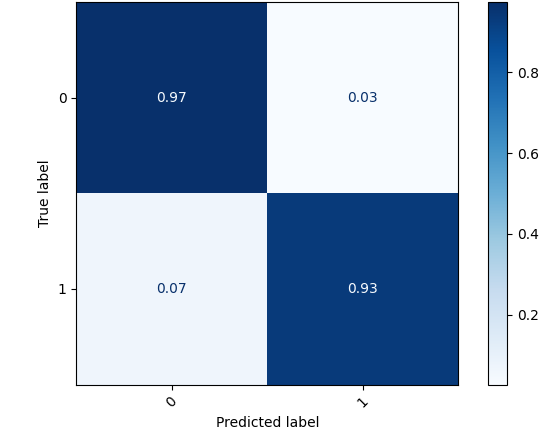}
    \caption{Confusion matrix of the classifier trained with the addition of the generalized Choquet-based feature for 4 features. }
    \label{XGConfMatrixChoquet}
\end{figure}

For Random Forest and XGBoost Without Choquet in Fig. \ref{XGConfMatrix}, the confusion matrix shows that the model correctly classifies $83\%$ of benign traffic (true label~0) and $99\%$ of attacks (true label~1), with $17\%$ of benign samples flagged as attacks (false positives) and only $1\%$ of attacks missed (false negatives). With the Choquet feature in Fig. \ref{XGConfMatrixChoquet},  benign detection improves substantially to $97\%$ (false positives drop to $3\%$), while attack detection decreases to $93\%$ (false negatives rise to $7\%$). This reflects a clear \emph{trade-off}: the Choquet aggregation greatly reduces false alarms on benign traffic, at the cost of a higher miss rate on attacks; the baseline does the opposite, identifying attacks extremely well but misclassifying more benign samples.

\subsection{Discussion of results}

Overall, the results consistently indicate that fuzzy aggregation via the generalized Choquet integral is most beneficial when the feature space is small and interactions among variables are not yet well represented by raw attributes. From a data perspective, the pipeline begins with a compact and reliable basis: the CIC-DDoS2019 dataset (692{,}703 rows, 79 columns) is cleaned and reduced through SelectKBest to a controlled search space of up to $k=10$ features, an upper bound chosen because the accuracy curves stabilize after 6--7 features for both models (Fig.~\ref{acRF} and \ref{acXG}). This design avoids injecting superfluous information while still exposing the performance trajectory as $k$ increases.

Methodologically, the comparison is fair: both pipelines (with and without the Choquet-based aggregated feature) share identical splits, training settings, and evaluation protocol.

Empirically, Fig.~\ref{acRF} and \ref{acXG} and Tables~\ref{tab:dffRF_swapped} and \ref{tab:dffXG} reveal a consistent pattern: with few features ($k!\in![1,5]$), the Choquet-augmented pipeline outperforms the baseline; as $k$ increases, the gap narrows, and both models converge for $k!\ge!6$--$7$. The largest improvement occurs at $k=4$, with $\Delta Y \approx 0.07$, indicating that fuzzy aggregation enables the classifiers to achieve high accuracy using fewer attributes.
The confusion matrices (Fig.~\ref{XGConfMatrix} and \ref{XGConfMatrixChoquet}, shown for the best-$k$ configuration) reinforce this result by highlighting a practical trade-off: the baseline favors maximum attack detection but produces more false positives on benign traffic, whereas the Choquet-based model substantially reduces false alarms (higher true-negative rate) at a modest cost in missed attacks.

In short, the generalized Choquet aggregation delivers better accuracy in low-dimensional data, enabling robust detection even in resource-constrained environments. Once the feature set surpasses $k\!\approx\!6$--$7$, the benefits decrease as both pipelines approach their ceiling. These findings support the use of Choquet-based fusion as a compact, interaction-aware representation that accelerates performance while controlling dimensionality.

Beyond the quantitative gains, the results highlight practical implications for large-scale network monitoring. The proposed Choquet-based aggregation offers a balanced trade-off between feature expressiveness and computational cost, making it suitable for deployment in data centers or backbone networks where massive traffic must be analyzed under strict latency constraints. Although the experiments were performed on offline batches of the CIC-DDoS2019 dataset, the methodology remains compatible with streaming frameworks, as the aggregation can be computed incrementally as new features arrive. This property indicates that the approach can be integrated into data analytics pipelines, enabling real-time anomaly detection without compromising interpretability or scalability.

\section{Conclusion}
This work presented a feature aggregation strategy based on a generalized Choquet integral to enhance network anomaly detection, particularly in scenarios with limited feature availability. By modeling nonlinear dependencies among selected attributes through fuzzy measures, the proposed approach improves predictive performance without requiring additional raw features.
An additional side effect of the experimental setup was a reduction in data volume, resulting from feature selection rather than the Choquet aggregation itself.

Furthermore, the results suggest a favorable trade-off: the Choquet model reduces false positives while slightly increasing false negatives, which can be balanced by threshold calibration. This high true-negative rate allows its use as an efficient first defense layer to filter benign traffic before more resource-intensive analysis. When evaluated over repeated stratified experiments, the performance gains were statistically significant ($p < 0.05$) for both classifiers, \red{positioning the generalized Choquet integral as a viable candidate for large-scale smart city deployments. In such scenarios, edge devices must process information locally to avoid saturating distributed cloud networks during massive DDoS events, making this high efficiency aggregation essential.}

Despite these gains, the approach has limitations. The aggregation's effectiveness remains dependent on the initial feature selection quality, and the computational complexity of calculating fuzzy measures increases with the number of features, which may hinder performance if $k$ is too large. Additionally, the current model requires careful threshold calibration to manage the observed trade-off between reducing false positives and the potential rise in false negatives.

Future work will focus on extending this framework to compare against alternative dimensionality reduction techniques such as PCA, autoencoders, and recursive feature elimination, as well as on analyzing the computational latency of the Choquet aggregation for real-time deployment. We also plan to conduct sensitivity analysis of the copula parameter $\alpha$ and the integration of the model with streaming-based intrusion detection systems.

\bibliographystyle{IEEEtran}
\bibliography{refs}

\vspace{12pt}

\end{document}